\documentclass[aps,prd,twocolumn,superscriptaddress]{revtex4}
 \usepackage[english]{babel} 
 \usepackage{latexsym} 
 \usepackage{amsmath} 
 \usepackage{amsfonts} 
 \usepackage{amssymb} 
 \usepackage{textcomp} 
 \usepackage{multirow}
 \usepackage{graphicx}  
 \usepackage{enumerate} 
 \usepackage{array} 
 \usepackage{tikz}
 \usetikzlibrary{fit,calc,positioning,decorations.pathreplacing,matrix}
 \usepackage[latin3]{inputenc} 
 \usepackage{graphics} 
 \usepackage{color} 
\usepackage{hyperref}

 \newcommand{\be}{\begin{equation}} 
 \newcommand{\ee}{\end{equation}} 
 \newcommand{\ba}{\begin{eqnarray}} 
 \newcommand{\ea}{\end{eqnarray}} 
 \newcommand{\ket}[1]{\arrowvert #1 \rangle}
 \newcommand{\bra}[1]{\langle #1 \arrowvert}
\begin{document} 
\title{Velocity fluctuations of fission fragments} 
\author{Felipe J. Llanes-Estrada, Bel\'en Mart\'{\i}nez Carmona and Jose L. Mu\~noz Mart\'{\i}nez}
\affiliation{Departamento de F\'{\i}sica Te\'orica I, Universidad Complutense de Madrid, Plaza de las Ciencias 1, 28040 Madrid, Spain.}
\begin{abstract}  
We propose event by event velocity fluctuations of nuclear fission fragments as an additional interesting observable that gives access to the nuclear temperature in an independent way from spectral measurements and relates the diffusion and friction coefficients for the relative fragment coordinate in Kramer-like models (in which some aspects of fission can be understood as the diffusion of a collective variable through a potential barrier). 
We point out that neutron emission by the heavy fragments can be treated in effective theory if corrections to the velocity distribution are needed.\\
$ $
\\
Preprint INT-PUB-15-070.
\end{abstract} 
\date{\today}

\maketitle 

\section{Introduction}

This brief report is concerned with a typical fission experiment in which two fragments 
are detected long after their scission and are later detected, having possibly emitted one or few neutrons along the way. 
We suggest that existing experiments track the event by event velocity fluctuations (or velocity variance) defined by
\begin{equation}
\label{sigma}
{{\sigma}_v}^2=\left< {v}^{2}\left(t\right)\right> -{\left< v\left(t\right)\right> }^{2}\ .
\end{equation}

Fission is a complex mechanism, and involves ``intrinsic'' degrees of freedom (individual or few-nucleon excitations in the fissioning nucleus) and collective degrees of freedom, such as the relative coordinate between the center of mass of the two major fragments ${\bf r}$.
The exchange of energy between these two types of degrees of freedom can be conveniently treated as a dissipative system (random walk on the ${\bf r}$ variable) and can be quantified by the dissipation coefficient $\gamma$ that will appear later in Eq.~(\ref{le}).

A picture of the process of historic interest is provided by the liquid drop model~\cite{Meitner,Bohr:1939ej}, followed shortly thereafter by the work of Kramers~\cite{Kramers:1940zz} who introduced a Langevin equation for the diffusion of the fission coordinate through a potential barrier.
The $O(200)$ MeV Coulomb barrier itself produces a large acceleration of the fission fragments, but this is a conservative $V(r)$ potential that produces a predictable velocity increase, unlike the stochastic forces that affect the fragments just before fission.

Within this conceptual framework, that has triggered theoretical work on and off for decades (e.g. \cite{Grange:1983zz}), 
 a salient concept is the Fluctuation-Dissipation Theorem (FDT). This is a relation between the attenuation of the motion of the interfragment ${\bf r}$ coordinate (induced for example by an incoming neutron) and the random fluctuations of the fragment velocities. A related quantity, the variance of the kinetic energy, was extensively studied in the past~\cite{Grossmann:1988}
Also studied in the past were fluctuations of the velocity as function of the mass number $A$~\cite{Straede}. This is distinct from the event by event velocity fluctuations that we propose.

We will consider these fluctuations with the three-dimensional Langevin approach and obtain a standard relation in statistical physics, this time between the coefficients of dissipation and diffusion (to our knowledge, this one so far unexplored).
Alternatively, the FDT can be used to obtain a relation between nuclear temperature and the velocity fluctuations, which is a possible different way of assessing nuclear temperature $T$ in the parent nucleus before fission.

\section{Three-dimensional Langevin equation and velocity fluctuations}

The three-dimensional Langevin equation (LE)~\cite{Aritomo:2014yia}
for ${v}_{i}=\dot{r}_i\equiv \frac{d{r}_{i}}{dt}$, with ${r}_{i}$ the components of the relative coordinate between the fragments, can be given as
\begin{equation}
\label{le}
\mu\dot{{v}_{i}}=-{\gamma}_{ij}{v}_{j}+{g}_{ij}{R}_{j}\left(t\right)+{F}_{i}\left({r}_{i}\right) \ .
\end{equation}
There, $\mu$ is the reduced mass of the two nuclear fragments; 
${\gamma}_{ij}$ is the friction tensor, whose coefficients are the dissipation coefficients; ${{R}_{i}(t)}$ is a stochastic force that we take normalized to Gaussian white noise, satisfying	
	\begin{equation}
	\label{R}
	\begin{split}
	\left< {R}_{i}\right> &=0\\
	\left< {R}_{i}(t){R}_{j}(t')\right> &=2{\delta}_{ij}\delta\left(t-t'\right)
	\end{split}  
	\end{equation}\\
where the ergodic hypothesis serves to compute the time average
 $\left< {A}\right> $ of any quantity $A$ as the average value over an ensemble.
The intensity of the stochastic force, ${g}_{ij}$, appearing in the FDT, 
is controlled by the ${D}_{ij}$ diffusion tensor,
	\begin{equation}
	\label{g}
	{g}_{ik}{g}_{jk}={D}_{ij}\ .
	\end{equation}\\
Finally, ${F}_{i}\left({r}_{i}\right)$ (the external force in Langevin's theory) is here a conservative force due to the potential barrier, 
$V$, that we take as time independent, ${ F }_{ i }\left( { r }_{ i } \right) =-{ \partial  }_{ i }V$.

We work with the force equation per unit mass by dividing 
 Eq.~(\ref{le}) by $\mu$, yielding
\begin{equation}
\label{LE}
\dot { { v }_{ i } } =-{ {\gamma}^*  }_{ ij }{ v }_{ j }+{ l }_{ i }\left( t \right) +{ f }_{ i }\left( { r }_{ i } \right) 
\end{equation}\\
where the dynamical quantities are now defined per unit mass
\begin{equation}
\label{def}
{ {\gamma}^*  }_{ ij }=\frac { 1 }{ \mu  } { \gamma  }_{ ij }\quad ,\qquad { l }_{ i }\left( t \right) =\frac { 1 }{ \mu  } { g }_{ ij }{ R }_{ j }\quad ,\qquad { f }_{ i }=\frac { 1 }{ \mu  } { F }_{ i }
\end{equation}

Due to the random force (the impulses caused by individual nucleons or few-nucleon clusters on the two separating fragments, the velocity derived from the relative coordinate fluctuates, and its fluctuations $\sigma_v^2$ defined in Eq.~(\ref{sigma})
are our major focuse in this work. 

These velocity fluctuations are, by fluctuation-dissipation relation (a manifestation of the FDT), proportional to a ratio of the diffusion tensor defined in Eq.~(\ref{g}) to the friction tensor of Eq.~(\ref{le}), 
\begin{equation}
\label{fluctuaciones}
\boxed{{{\sigma}_v}^2=\frac{1}{\mu}Tr\left(D{\gamma}^{-1}\right)} \ .
\end{equation}\\
Its derivation is technical but standard and we relegate it to appendix~\ref{app:deriveFDT}. In the one-dimensional case one simply has
\begin{equation}
\label{fluc1}
{{\sigma}_v}^2=\frac{D}{\mu\gamma},
\end{equation}

A key remark is that in Eq.~(\ref{fluctuaciones})
the contributions of the large barrier force ${F}_i$ have cancelled. This means that, even there being a large energy given to the fragments by the Coulomb barrier, this does not manifest itself in their velocity fluctuations, that are purely due to other stochastic processes.

\subsection{Assessing nuclear temperature from $\sigma_v$}
Excited nuclei can be assigned an approximate temperature controlling the occupancy of the various energy levels. This requires them to be in a state of equilibrium or nearly so, and it is evidenced for example in fits to gamma radiation or neutron spectra~\cite{povh} that follow a Maxwell distribution
\begin{equation}
\label{maxwell}
{N}_{n}\left({E}_{n}\right)\sim\sqrt{{E}_{n}}\cdot{e}^{\frac{-{E}_{n}}{kT}}\ .
\end{equation}

Eq.~(\ref{fluctuaciones}) above is a particular expression of the Fluctuation-Dissipation theorem relating the system response to an external perturbation to the fluctuations in thermal equilibrium, and Einstein's relation applies~\cite{frobrich},
\begin{equation}
\label{eins}
{D}_{ij}={\gamma}_{ij}T\ .
\end{equation}\\
If we employ Eq.~(\ref{eins}) in Eq.~(\ref{fluctuaciones}), we obtain
\begin{equation}
\label{T}
{{\sigma}_v}^2=\frac{T}{\mu}
\end{equation}\\
so the velocity fluctuations provide a method to obtain the nuclear temperature that is alternative to Eq.~(\ref{eins}).

Moreover, Eq.~(\ref{T}) allows an estimate of the order of magnitude of the velocity fluctuations, so we get an idea of what precision
in fragment velocity measurements is necessary to access intrinsic (as opposed to detection or instrumental) nuclear fluctuations.

Let us take as simple example the two-fragment asymmetric fission of ${}^{252}$Cf, with temperature about $\sim1.4$MeV. 
Setting {\it e.g.} $m_1=\frac{M}{3}$ and $m_2=\frac{2M}{3}$, 
$[\mu_{\rm red }\simeq 55 \text{ GeV}]$,
\begin{equation}
\label{estimacionsigma}
{{\sigma}_v}^2=\left< v^2\right> -{\left< v\right> }^2 \sim 2.5\times {10}^{-5}\ .
\end{equation}
If the index $i=1,..., N$ swipes all collisions of an experimental run,
$\sum_{i}{\frac{{v_i}^2}{N}}-{\left(\sum_i{\frac{v_i}{N}}\right)}^2\sim2.5\times {10}^{-5}\rightarrow\sum_{i}{{v_i}^2}-\frac{1}{N}{\left(\sum_i{v_i}\right)}^2\sim
2.5\times {10}^{-5}N]$.
We can propagate the error of the experimental measurement $v_i\longrightarrow v_i+\Delta v_i$, absent systematic shift, $\left< \Delta v\right>=0$, to 
$[\left(\left< v^2\right>-{\left< v \right> }^2\right)+2\left< v\Delta v\right> \sim 2.5\times {10}^{-5}]$. The term $2\left< v\Delta v\right>  $ dominates and must not overwhelm the experimental signal, so we need to request from experiment that
\begin{equation}
\label{errv}
2\left< v\Delta v\right>  < 2.5\times {10}^{-5}
\end{equation}
where, again, the average is taken over many collisions and $v\Delta v$ is the projection of the error over the actual velocity, so we express it more conveniently as an error in the fragment's kinetic energy $T_{KE}=\frac{1}{2}\mu v^2$, and
therefore, $\Delta T_{KE}\simeq2.5\times \frac{{10}^{-5}}{2}\cdot55$ GeV $\sim0.7$ MeV. As the typical kinetic energy is in the range $50-100$ MeV, 
\begin{equation}
\label{errK}
\frac{\Delta T_{KE}}{T_{KE}}\sim\left(0.7-1.4\right)\times{10}^{-2}
\end{equation}\\
it is enough to achieve a precision of $0.5\%$ in measuring fragment kinetic energy to access the intrinsic fluctuations.

\subsection{Quantum fluctuations}
Inasmuch as the nucleus is not excessively hot, we near a degenerate-fermion system, and quantum fluctuations may be a concern, with a velocity variance $\sigma_v$ appearing even for near $T=0$ fissioning nuclei,
$ {{\sigma}_v}^2(T=0) = {\left({{\sigma}_v}^2\right)}_{q}  $

One can define~\cite{Roscilde} an effective quantum-fluctuation pseudotemperature $T_{\rm eff}$ through ${\left({{\sigma}_v}^2\right)}_{q}=\frac{{T}_{\rm eff}}{\mu}$  
but it is perhaps clearer to use the quantum formulae directly.
For a quantum oscillator $H=\frac{p^2}{2\mu}+\frac{kx^2}{2} = \hbar \omega_0 (c^\dagger c +\frac{1}{2})$, with thermal occupation number (computed as usual from $\left< A\right> = {\rm Tr}[Ae^{-\beta H}]$)
\begin{eqnarray}
\left< c^\dagger c\right> &=& n(\omega_0) = \frac{1}{e^{\beta\hbar \omega_0}-1} \nonumber \\
&=& \frac{1}{2} \coth\left(\frac{\beta\hbar\omega_0}{2}\right)\ ,
\end{eqnarray}
one obtains
\begin{equation}
\left< x(t)x(0) \right> = \frac{\hbar}{2\mu k} \left(
\coth \frac{\beta \hbar \omega_0}{2} \cos (\omega_0 t) -i \sin (\omega_0 t)
\right)
\end{equation}
and from
\begin{equation}
\left< p(t) p(0) \right> = -\mu^2 \frac{d^2 \left< x(t)x(0)\right>}{dt^2}
\end{equation}
finally
\begin{equation} \label{Tquant}
\left< v^2 \right>_q = \frac{\hbar}{2} \sqrt{\frac{k}{\mu^3}} \coth \frac{\beta \hbar \omega_0}{2}
\end{equation}
that in the $T\to 0$ limit reduces to 
\begin{equation}  \label{quantfluct}
\left< v^2 \right>_q = \frac{\hbar}{2} \sqrt{\frac{k}{\mu^3}} \ .
\end{equation}

Since this oscillator is centered at 0, Eq.~(\ref{quantfluct}) is directly $(\sigma_v^2)_q$ which is thus interpreted
in terms of its spring constant. The quantum versus classical nature of the fluctuations can be separated from experimental data by 
comparing the two temperature dependences in Eq.~(\ref{T}) (the classical expression) and Eq.~(\ref{Tquant})  (the quantum expression).
It would be interesting to experimentally see in this particular observable the onset of the quantum regime at low temperature (in practice one would use the nuclear excitation energy as a proxy for the temperature, along the lines of $T=\sqrt{8{\rm MeV} \times E/A}$ ). Obviously, fluctuations of the velocity remain at very low
temperature and they might be extracted.

\section{Extraction of the diffusion coefficient}

Equation~(\ref{fluc1}) above is a simple gauge to measure the diffusion coefficient given knowledge of the dissipation one,
that has to be known from other sources.

A possible way to measure the later is through a hydrodynamic fit to fission, path followed {\it e.g.} by some GANIL experiments\cite{sanchez}. 
There, a ${}^{208}$Pb beam collides with liquid Hydrogen so proton-induced fission of lead results. Fokker-Planck simulations of the nuclei are run for different values of the reduced dissipation coefficient ($\beta$ in that work, corresponding to our ${\gamma}^*$). The fitted value is within the expected nuclear scale, $\gamma^*=4.5\cdot{10}^{21}\text{ s}^{-1}\simeq2.96\text{ MeV}$. Earlier theoretical predictions~\cite{Grossmann:1988} were in the range 0.92 (for $^{238}$U) to 1.71 (for $^{238}$Fm) $\times 10^{21}$s$^{-1}$.
 
Either of Eq.~(\ref{fluc1}) or Einstein's relation in Eq.~(\ref{eins})
then gives us the diffusion coefficient $D$,
\begin{equation}
\label{D2}
D=\mu {\gamma}^* T \ .
\end{equation}
With our typical estimate for ${{\sigma}_{v}}^{2}$ in Eq.~(\ref{estimacionsigma}) we can then quote
\begin{equation}
D
\simeq  225\text{MeV}^3\ .
\end{equation}

What we advocate here is to use Eq.~(\ref{fluc1}) together with a measurement of $\sigma_v^2$ and 
separate experimental extraction of the dissipation coefficient $\gamma$ for a given fragment reduced mass $\mu$,
to finally yield the diffusion coefficient $D$. Note that this is the \emph{momentum space} diffusion coefficient, 
with the position space coefficient being related to it by Fokker-Planck theory (see the appendix of~\cite{Abreu:2011ic}), 
that yields $D_x = \frac{T^2}{D}$. Fokker-Planck theory also relates these coefficients to the scattering rate of the 
fragment-fragment relative particle with the medium nucleons, so their dynamical content can be eventually related to more fundamental
quantities.

In the next section we proceed to address the fragment velocity since it is the main element that our new observable calls for.
\section{Fragment velocity}

The typical average velocity of the fragments can be estimated from the available energy and mass-fragment distribution. Experimental determination of the nuclear mass is so accurate that the effect of $\Delta M$ is negligible 
against the uncertainty in the measurement of the kinetic energy $\Delta \left<T_{KE}\right>$, such that
\begin{equation}
\frac{\Delta \left<v\right>}{\left<v\right>}\simeq\frac{\Delta \left<T_{KE}\right>}{2\left<T_{KE}\right>} \ .
\end{equation}

Bertsch {\it et al.}~\cite{benchmarking}, collect data for thermal-neutron induced ${}^{235}$U, and ${}^{239}$Pu fission, and spontaneous ${}^{252}$Cf fission that we can use as an example.

\begin{figure}[hbtp]
\centering
\includegraphics[width=0.48\textwidth]{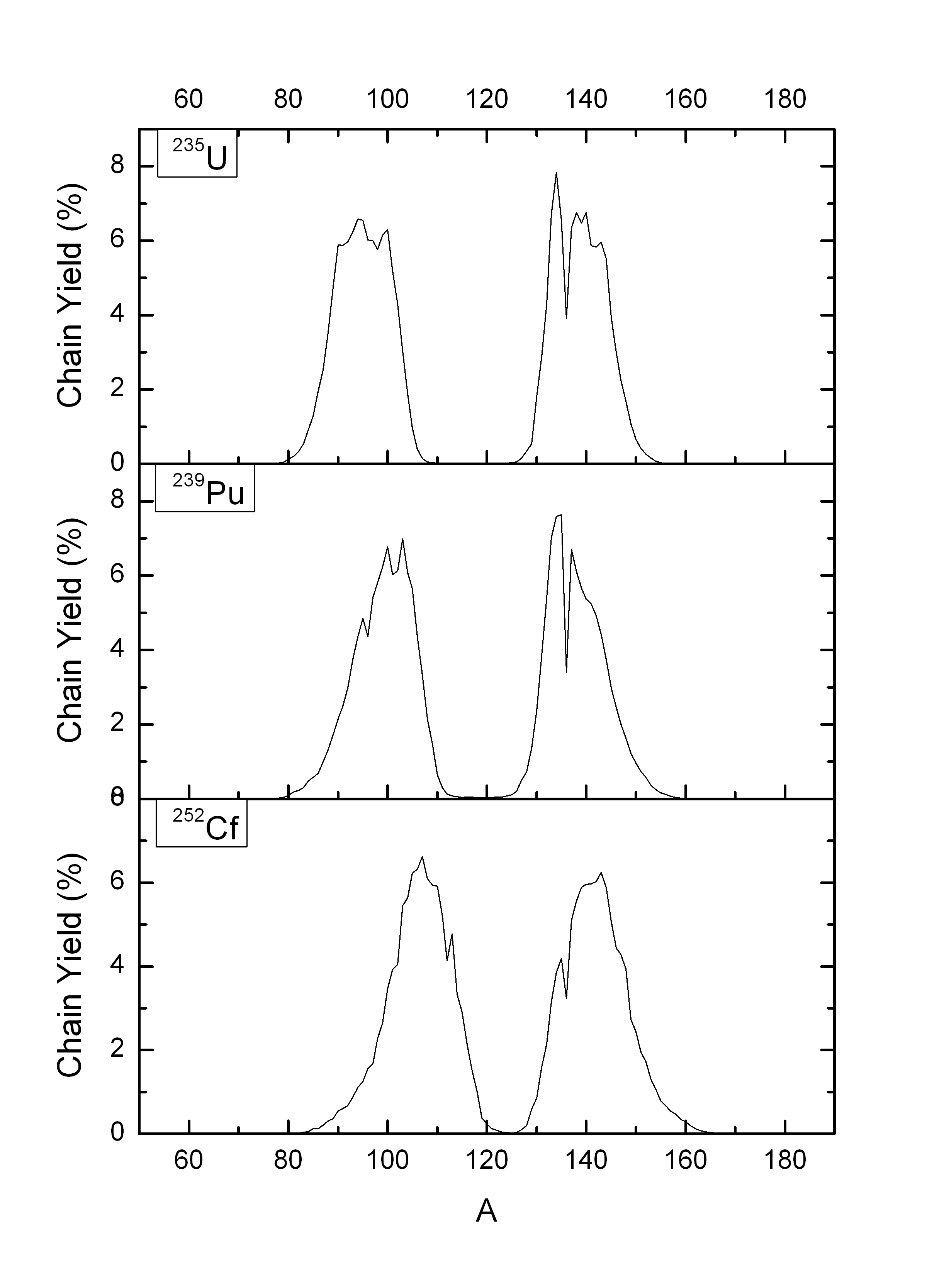}
\includegraphics[width=0.48\textwidth]{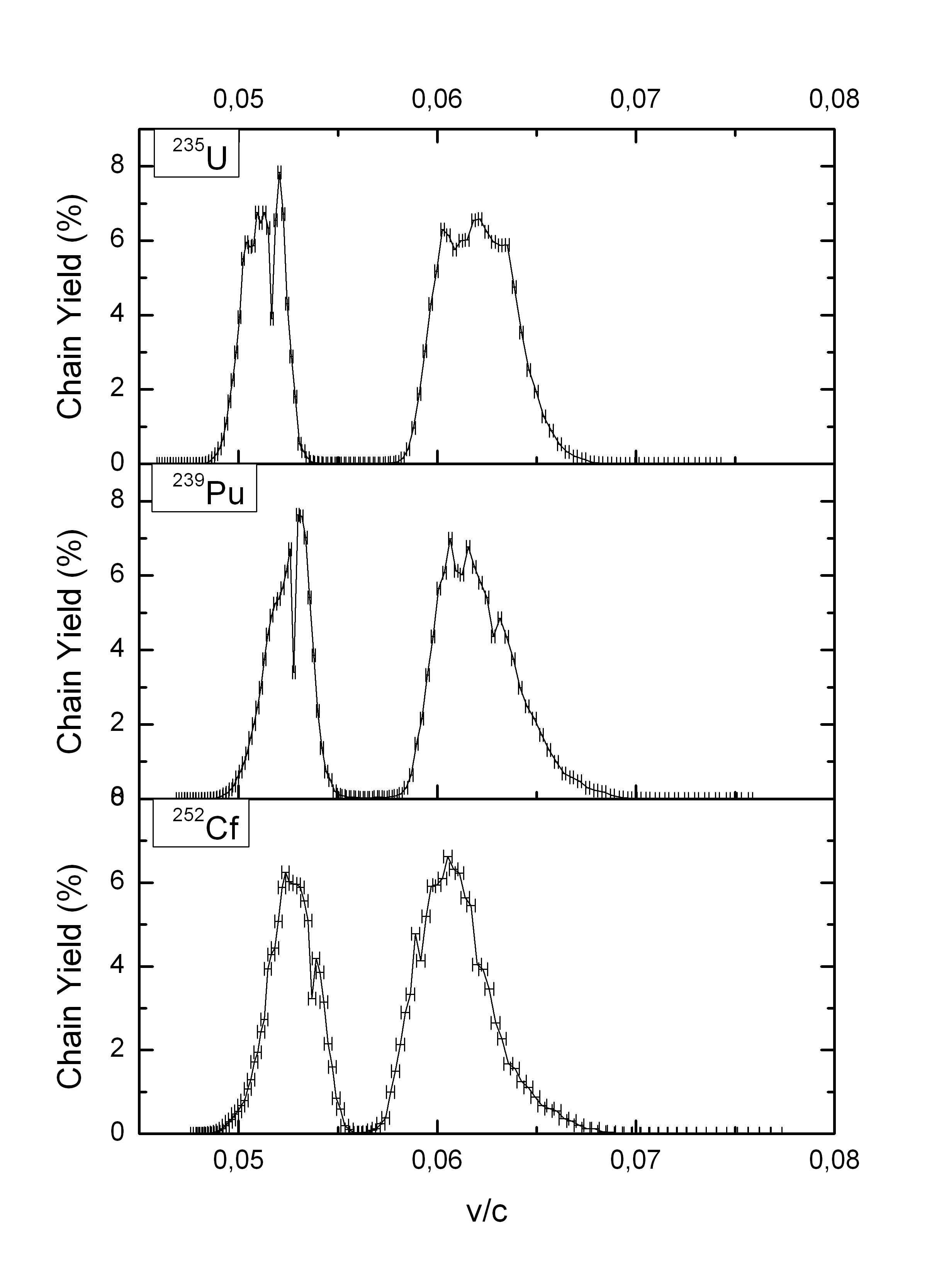}
\caption{{\small 
Rendering of the fission mass fragment distribution (top) for
${}^{235}$U, ${}^{239}$Pu and ${}^{252}$Cf~\cite{benchmarking} and rough estimate of the velocity distribution (bottom).}}
\label{distrA}
\end{figure}

\begin{figure}
\includegraphics[width=0.9\columnwidth]{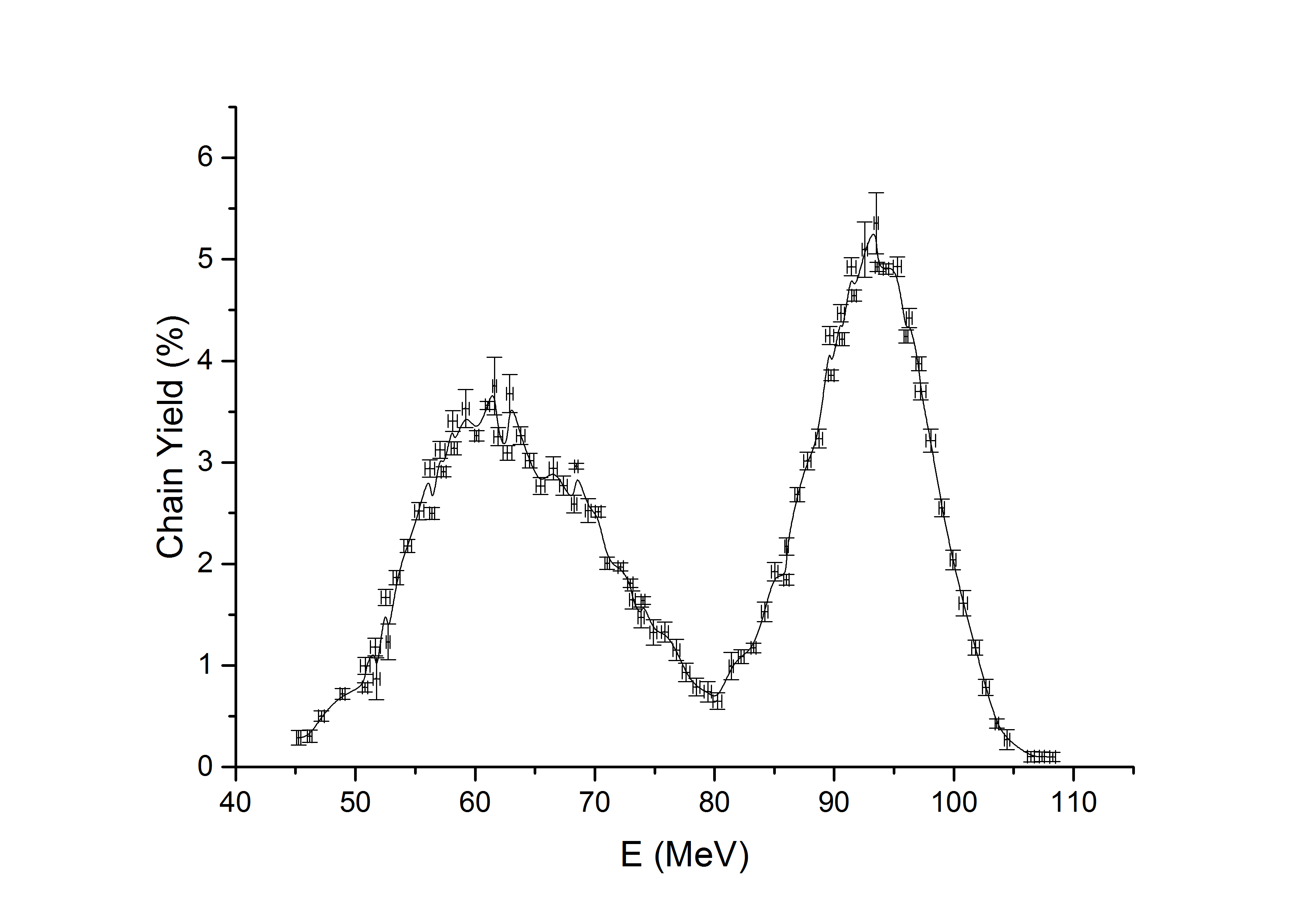}
\caption{\label{fig:distE} Average energy for the fission fragments of $^{235}U$ fission as function of fragment $A$.}
\end{figure}

Fig. \ref{distrA} shows those asymmetric mass distributions. With a crude, illustrative fragment-mass estimate for $M=\bar{Z}{m}_{p}+\left(A-\bar{Z}\right){m}_{n}-8{\rm MeV}A$ and nonrelativistic kinematics
\begin{equation}
\begin{split}
\left<v\right>&=\sqrt{\frac{2\left<T_{KE}\right>}{M}}\\
\Delta \left<v\right>&=\sqrt{{\left(\frac{\left<v\right>}{2}\frac{\Delta \left<T_{KE}\right>}{\left<T_{KE}\right>}\right)}^2+{\left(\frac{\left<v\right>}{2}\frac{\Delta M}{M}\right)}^2}
\end{split}
\end{equation}\\
(where $\left<T_{KE}\right>$ is the fragment's average kinetic energy, that turns out to be $170.5\pm0.5$ for ${}^{235}$U ($n_{th}$,$f$), $177.9\pm0.5$ for ${}^{239}$Pu ($n_{th}$,$f$) and $184.1\pm1.3$ for ${}^{252}$Cf in spontaneous fission).
This is confirmed by the direct measurements~\cite{Fowler} of the energy distribution of the fission fragments, such as shown in figure~\ref{fig:distE}. The average energy for all fragments is in this case 156 MeV, with modal peaks for light and heavy fragments at about 93.5 and 61.6 MeV respectively.

Exemplifying with ${}^{239}\text{Pu}+n\longrightarrow\text{Zr}+\text{Cs}$, 
$\frac{\Delta\left<v\right>}{\left<v\right>}\sim0.0014\longrightarrow\frac{\Delta T_{KE}}{T_{KE}}\sim2.8\cdot10^{-3}$.
Comparing them with equation~(\ref{errK}), we find that contemporary experimental measurements are more than sufficiently precise so as to extract ${\sigma_v}^2$. 
For example, \cite{Caamano:2015sva} provides a number of new observables and particularly measurements of fission fragment velocity distributions in the fission of $^{240}$Pu and $^{250}$Cf
that fall almost linearly between 1.5 and 0.9 cm/s
with the fragment's $Z\in(35,65)$.

\subsection{Neutron emission does not alter \emph{velocity} distributions}
We have consistently discussed \emph{velocity} as opposed to \emph{momentum} distributions, and in this subsection we clarify why.
The reason is that all quantities are measured at asymptotically large times at detectors, but one would like to know the fluctuations at the instant of fission. However, the fragments often lose energy in fly, saliently by neutron emission. 
These neutrons are indeed emitted with average number $2.48$ for $^{233}U$, $2.42$ for $^{235}U$ and $2.86$ for $^{239}Pu$, and (as well as any other radiation), alter the momentum distribution.

But fragment velocities $v_i=p/M_i$ barely change. Moreover, an effective theory can be formulated, analogous to Heavy Quark Effective Field Theory~\cite{Georgi} with the nuclear fragment playing the role of the heavy quark, and the emitted neutron or other radiation, that of light quarks and gluons.

We exemplify the ideas with two spinless fragment $\phi_{1}$ and $\phi_{2}$ fission, followed by emission of two neutrons~\footnote{For instance $^{236}U^*\sim n + ^{235}U \longrightarrow n\; + n\; +\; ^{A}X\; +\; ^{(236-A-2)}Y$, with both fragments being even.}, but generalization is immediate.

A simple Lagrangian density $\mathcal{L}= \mathcal{L}_{\rm free} + \mathcal{L}_{\rm int}$ reflecting the neutron emission can be taken as

\begin{eqnarray}
\label{Lagrangian2n}
\mathcal{L}_{\rm free}=\frac{1}{2}\partial_{\mu}\phi_{1}^{\dagger}\partial^{\mu}\phi_{1}-m_{1}^{2}\phi_{1}^{\dagger}\phi_{1} 
\nonumber \\ \nonumber
+ \frac{1}{2}\partial_{\mu}\phi_{2}^{\dagger}\partial^{\mu}\phi_{2}-m_{2}^{2}\phi_{2}^{\dagger}\phi_{2}  
\\ \nonumber
\mathcal{L}_{\rm int}=\int_{xy}\phi_{2'}^{\dagger}(x)\phi_{2'}(x) V(\vert x-y\vert) \phi_{1}^{\dagger}(y)\phi_{1}(y) 
\\
+ \int_{x} \; dx\; g\; (\overline{n^{c}}n) \phi_{2'}^{\dagger} \phi_{2} + \dots
\end{eqnarray}

The free Lagrangian provides the Klein-Gordon equation for each fragment. The interacting one has a part stemming from the interfragment potential, 
which is related to the acceleration, 
a piece related to the production of neutrons with spins $\sigma$ and $\tau$ and further terms that do not impact our arguments.
The field $\phi^{\dagger}_{2'}$ refers to the second fragment after emitting the two neutrons (the first fragment does not evaporate any nucleons in this example).
The stochastic force of Eq.~(\ref{le}) does not appear in the Lagrangian density of Eq.~(\ref{Lagrangian2n}) because it describes the situation after scission. Those interactions are therefore free of dissipation.

The fields  $\phi_{i}(y)$,  ($i=1,2$) and $n(x)$ may be expanded in their normal modes
\begin{eqnarray}
\phi_{i}(y)= \int^{\Lambda} d^{3} q \left[ a_{i}(q)e^{iqy} + b_{i}^{\dagger}(q)e^{-iqy}\right] 
\\
n(x)=\sum_\tau\int^{\Lambda} d^{3} q \left[ n_{\tau}(q)u_{\tau}(q) e^{iqx} + n_{\tau}^{\dagger}(-q)v_{\tau}(k) e^{-iqx}\right] \ .
\end{eqnarray}
As the nuclear fragments are the heavy degrees of freedom, we may ignore the antiparticle terms, suppressed by  $\frac{E}{2M_{i}}$.

Near the scission point, only the two fragments are present with momenta $k$ and $-k$. After Coulomb-induced separation and neutron emission (double, in this example), total momentum is composed of the fragments $k_{1}$ and $k_{2}$, and of the neutron $k_{3}$ and $k_{4}$ momenta. Thus, the 
initial (before neutron emission) $\ket{i}$ and final $\ket{f}$ states
may be written as
\begin{equation}
\ket{i}= a_{1}^{\dagger}(k) a_{2}^{\dagger}(-k) \ket{0}
\end{equation}
\begin{equation}
\ket{f}=a_{1}^{\dagger}(k_{1}) a_{2'}^{\dagger}(k_{2}) n_{s_3}^{\dagger}(k_{3}) n_{s_4}^{\dagger}(k_{4}) \ket{0}
\end{equation}

The matrix element for the two-neutron emission does not depend
on the interfragment potential $V$, 
\begin{eqnarray} \label{matrixel}
\bra{i}\mathcal{L}_{int}\ket{f}
= \int dx\bra{i}   g\; (\overline{n_{\sigma}^{c}}n_{\tau}) \phi_{2'}^{\dagger} \phi_{2}\ket{f}\ .
\end{eqnarray}

\begin{figure}[h]
\includegraphics[width=6cm]{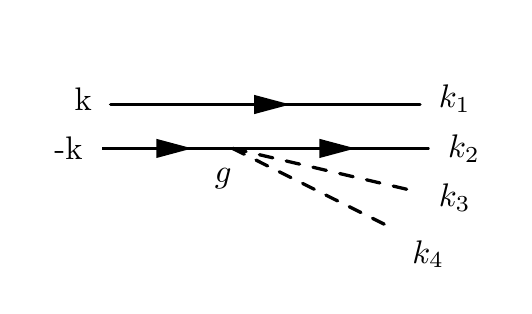}
\caption{Feynman diagram for fragment 2 emitting a neutron pair (1 neutron is not possible as all nuclear fragments are taken as scalars). }
\label{diagramasfey}
\end{figure}

The Feynman diagrams associated to Eq.~(\ref{matrixel}) are the one shown in figure~\ref{diagramasfey} and another with the two identical neutrons exchanged.

Operating in Eq.~(\ref{matrixel}),
\begin{equation}
\begin{split}
\bra{i}\mathcal{L}_{int}\ket{f}= g \sum_{\sigma, \tau} (u_{\sigma}^{T}Cu_{\tau}) \\
\delta (k-k_{1}) \delta(k+k_{2}+k_{3}+k_{4})(\delta_{s_{3}\tau}\delta_{s_{4}\sigma}-\delta_{s_{3}\sigma}\delta_{s_{4}\tau})
\end{split}
\end{equation}
we recognize typical $nn$ spin-zero emission. This heavy-fragment effective theory offers a systematic way of classifying $1/M$ suppressed corrections, but for our leading-order argument all we need is to track the kinematics, contained in the conservation $\delta$s, 
\begin{equation}
\label{DD}
\bra{i}\mathcal{L}_{int}\ket{f}\propto \delta (k-k_{1}) \delta(k+k_{2}+k_{3}+ k_{4})\ .
\end{equation}
Converting to velocities with $k_{i}=m_{i}v_{i}$, Eq.~(\ref{DD}) yields
\begin{equation}
\bra{i}\mathcal{L}_{int}\ket{f}\propto  \delta (k-k_{1})\;m_{2}\;\delta\left( (v+v_{2})+\frac{k_{3}+k_{4}}{m_{2}}\right) \ ;
\end{equation}
as fragment masses are of $O$(GeV) and neutron momentum $O$(MeV),
we see that the velocity of the nuclear fragments in the initial state is not changed by the radiation. Therefore, the recoil velocity at the instant of scission is directly measured in the final state up to $E/M$ corrections, $\vert v\vert=\vert v_{2}\vert+O\left( \frac{k_{3}+k_{4}}{m_{2}}\right)$.

\begin{figure}[h]
\centering
\includegraphics[scale=0.5]{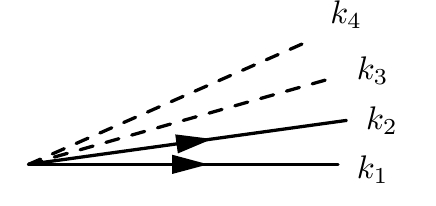}
\caption{A further example diagram for double neutron emission (prompt emission simultaneous to fission).  \label{fig:further} }
\end{figure}

The example Lagrangian above does not include other possible dynamical terms such as the diagram in fig.~(\ref{fig:further}) involving prompt emission simultaneous to fission, nor radiation with intermediate spin 1/2 nuclei, etc.; but no specific dynamics can change the simple kinematic counting, and thus we believe the velocities are measurable to an error of order $10^{-4}$.

In the end, the error in the measurement of $\sigma_v^2$ induced by the emission of neutrons with joint momentum $k$ is
\begin{equation}\label{corrections}
\Delta \sigma_v^2|_{\rm n\ emission} = \frac{2}{M} \left(
\langle vk \rangle -\langle v\rangle \langle k \rangle \right) + O\left(M^2 \right)\ .
\end{equation}
This equation means that for given, fixed $k$, the fluctuations are not affected at order $1/M$. Thus, $\Delta \sigma_v^2$ is suppressed by $1/M$ respect to $\sigma_v^2$,
and in any case the two numbers $k$ and $v$ are decorrelated (since the neutron momentum has to do with the intrinsic temperature of the fragment
evaporating it whereas the fragment velocity is related to the fluctuations during scission time) and therefore the $O(1/M)$ term in Eq.~(\ref{corrections}) 
vanishes.

\section{Discussion}

Past work on fluctuations in fission focused very much on kinetic energy fluctuations. We have pointed out that actually velocity fluctuations carry very direct information about what was the situation at the scission point because any radiation (for example, in the form of neutrons) carries energy and momentum away from the nuclear fragments, but it barely alters their velocity, which is therefore a ``relic'' of earlier fission stages. In principle one can construct specific Heavy Fragment Effective Theories for reactions of most interest, though this will be labor intensive as many channels with different spins would need to be described. Nevertheless, we find this is a tool to organize thought and expose the validity of the classical leading order that ignores $1/M$ corrections.

Our prediction of the independence of the velocity fluctuations $\sigma_v$ from neutron evaporation is testable by looking at the difference of same-$Z$ velocity with varying $A$, which recent experiments\cite{Caamano:2015sva} show as possible by identifying the neutron excess in each fragment.
The techniques in that work allow the reconstruction of the velocities at scission.

As for the experimental extraction of the $\sigma_v$ variance, one just needs to recall that time averages are of course, thanks to the ergodic hypothesis, 
obtained from event averaging, for example
\begin{equation}
\sigma_v^2=\frac{1}{N}\left(\sum_iv_i^2-\frac{\left(\sum_iv_i\right)^2}{N}\right) \ .
\end{equation}
In fact, even before the experimental extraction of the event-by-event fluctuations, a Montecarlo extraction might be useful to further characterize what elements of the theoretical description of the fission process affect these fluctuations. There are groups that are in a situation to attempt such a simulation~\cite{Vogt}.

As a curiosity, we mention that the Lorentz-invariant operator for two-neutron emission $\overline{n_{\sigma}^{c}}n_{\tau}$ in Eq.~(\ref{matrixel}) is familiar from neutron-antineutron oscillation theory
~\cite{Gardner:2014cma,Berezhiani:2015uya}. There of course, the operator is used by itself and the Hamiltonian violates baryon number conservation in two units, here,
as appropriate for strong-interaction theory, baryon number is compensated by $\phi_{2'}^{\dagger} \phi_{2}$ (the two fields representing different isotopes) and thus conserved in the process.

\section*{Acknowledgments}
The work of FJLE relied on the Spanish Excellence Network on Hadronic Physics FIS2014-57026-REDT, and grants UCM:910309, MINECO:FPA2014-53375-C2-1-P. He also thanks
the Department of Energy's Institute for Nuclear Theory at the University of Washington for its partial support and hospitality, particularly George Bertsch for advising on the manuscript.

\appendix

\section{Derivation of the Fluctuation-Dissipation relation}
\label{app:deriveFDT}

We now obtain Eq.~(\ref{fluctuaciones}). For this, we will formally solve Eq.~(\ref{LE}) and calculate the average values of ${v}_i(t)$ and of ${v}^2(t)$.

The general first integral of Eq.~(\ref{LE}) with initial value 
${\bf v }_{ o } $ is of the form
\begin{eqnarray}
\label{v}
{ v }_{ i }\left( t \right) &=&{ \left( { e }^{ -{ {\gamma}^*  }t } \right)  }_{ ij } { { v }_{ o } }_{ j } \\ \nonumber  
&+& { \left( { e }^{ -{ {\gamma}^*  }t } \right)  }_{ ij } \int _{ 0 }^{ t }{ { \left( { e }^{ { {\gamma}^*  }t' } \right)  }_{ jk } } \left[ { l }_{ k }\left( t' \right) +{ f }_{ k }\left( { r' }_{ k } \right)  \right] dt'\ .
\end{eqnarray}

We calculate its average value, taking into account  
that the stochastic force ${l}_{i}$ satisfies Eq.~(\ref{R}) in the second term and noting that the average of a constant is that same constant in the first one, to obtain
\begin{eqnarray}
\label{<v>}
{ \left<  v\left( t \right)  \right>   }_{ i }&=&{ \left( { e }^{ -{\gamma}^* t } \right)  }_{ ij } { { v }_{ 0 } }_{ j } \\ \nonumber
&+&{ \left( { e }^{ -{\gamma}^* t } \right)  }_{ ij } \int _{ 0 }^{ t }{ { \left( { e }^{ {\gamma}^* t' } \right)  }_{ jk }{ \left<  f\left( { r' }_{ k } \right)  \right>   }_{ k }dt' } \ .
\end{eqnarray}
We next obtain ${ \left<  v\left( t \right)  \right>   }^{ 2 }$ 
by left-multiplying Eq.~(\ref{<v>}) with its transpose,
\begin{equation}
\label{<v>2}
\begin{split}
{ \left<  v\left( t \right)  \right>   }^{ 2 }&={ { \left<  v\left( t \right)  \right>   }^{ T } }_{ i }{ \left<  v\left( t \right)  \right>   }_{ i }\\
&={ { v }_{ 0 } }_{ j }{ \left( { e }^{ -2{\gamma}^* t } \right)  }_{ jl }{ { v }_{ 0 } }_{ l } \\
&+{ { v }_{ 0 } }_{ j }{ \left( { e }^{ -2{\gamma}^* t } \right)  }_{ jl }\int _{ 0 }^{ t' }{ { \left( { e }^{ {\gamma}^* t'' } \right)  }_{ lm }{ \left<  f\left( { r'' }_{ m } \right)  \right>   }_{ m }dt'' } \\
&+\int _{ 0 }^{ t }{ { \left<  f\left( { r' }_{ k } \right)  \right>   }_{ k }{ \left( { e }^{ {\gamma}^* t' } \right)  }_{ kj }dt' } { \left( { e }^{ -2{\gamma}^* t } \right)  }_{ jl }{ { v }_{ 0 } }_{ l } \\
&+\!\!\int_0^t \!\! \int_0^{t'}\!\!\!\left<  f_k(r'_k) f_m(r^{''}_m)  \right>  \left(e^{\gamma^* t'}e^{-2\gamma^* t} e^{\gamma^* t''} \right)_{km}dt'dt'' 
\end{split} 
\end{equation}
Following a similar procedure from Eq.~(\ref{v}) we can then calculate $v^2(t)$ 
\begin{equation}
\label{v2}
\begin{split}
{ v }^{ 2 }\left( t \right)&={ \left( { v }_{ i }\left( t \right)  \right)  }^{ T }\left( { v }_{ i }\left( t \right)  \right)\\
&={ { { v }_{ 0 } }_{ j } }{ \left( { e }^{ -2{\gamma}^* t } \right)  }_{ jl }{ { v }_{ 0 } }_{ l } \\
&+{ { { v }_{ 0 } }_{ j } }{ \left( { e }^{ -2{\gamma}^* t } \right)  }_{ jl }\int _{ 0 }^{ t' }{ { \left( { e }^{ {\gamma}^* t'' } \right)  }_{ lm }\left[ { l }_{ m }\left( t'' \right) +{ f }_{ m }\left( { r'' }_{ m } \right)  \right] dt'' } \\
&+\int _{ 0 }^{ t }{ \left[ { l }_{ k }\left( t' \right) +{ f }_{ k }\left( { r' }_{ k } \right)  \right] { \left( { e }^{ {\gamma}^* t' } \right)  }_{ kj }dt' } { \left( { e }^{ -2{\gamma}^* t } \right)  }_{ jl }{ { v }_{ 0 } }_{ l } \\
&+\!\int_0^t \! \int_0^{t'}\!\left[l_k(t') +f_k(r'_k)\right]
\left(e^{\gamma^* t'} e^{ -2\gamma^* t}e^{\gamma^* t''}\right)_{km}
\\ & \phantom{+\int\int}
\left[l_m(t'')+f_m(r''_m)  \right]\! dt'dt'' 
\end{split}
\end{equation}
and $\left< v^2(t)\right> $ 
\begin{equation}
\label{<v2>}
\begin{split}
\left<  { v }^{ 2 }\left( t \right)  \right>  &={ { { v }_{ 0 } }_{ j } }{ \left( { e }^{ -2{\gamma}^* t } \right)  }_{ jl }{ { v }_{ 0 } }_{ l }\\
&+{ { { v }_{ 0 } }_{ j } }{ \left( { e }^{ -2{\gamma}^* t } \right)  }_{ jl }\int _{ 0 }^{ t' }{ { \left( { e }^{ {\gamma}^* t'' } \right)  }_{ lm }{ \left<  f\left( { r'' }_{ m } \right)  \right>   }_{ m }dt'' } \\
&+\int _{ 0 }^{ t }{ { \left<  f\left( { r' }_{ k } \right)  \right>   }_{ k }{ \left( { e }^{ {\gamma}^* t' } \right)  }_{ kj }dt' } { \left( { e }^{ -2{\gamma}^* t } \right)  }_{ jl }{ { v }_{ 0 } }_{ l }\\
&+\!\!\int_0^t\!\! \int_0^{t'}\!\!\! \left<  \left[l_k(t')\!+\! f_k(r'_k)  \right]\left(e^{\gamma^* t'} e^{-2\gamma^* t } e^{\gamma^* t'' } \right)_{km} \right. \\ & \phantom{+\int \int} \left.
\left[l_m(t'') +f_m(r''_m)  \right]  \right>  dt'dt''
\end{split}
\end{equation}

Let us delve a moment on the last term of Eq.~(\ref{<v2>}). 
A small manipulation and use of Eq.~(\ref{R}) gives
\begin{equation}
\label{termino}
\begin{split}
&\int_0^t\int_0^{ t' }\left<  \left[l_k(t') +f_k(r'_k)  \right]
\left(e^{\gamma^* t' } e^{ -2\gamma^* t}  e^{\gamma^* t'' } \right)_{ km }
\right.
\\ &  \left. \phantom{ e^{\gamma^* t'' }}
\left[ l_m(t'') + f_m(r''_m)  \right]  \right>  dt'dt''  \\
&=\int_0^t \int_0^{t'} \left<   f_k(r'_k) f_m(r''_m)  \right> 
 \left( e^{\gamma^* t' } e^{ -2\gamma^* t} e^{\gamma^* t'' } \right)_{km}dt'dt'' \\
&+\int_0^t\int_0^{t'}\left<  l_k(t') l_m(t'')  \right>  
 \left( e^{ \gamma^* t' } e^{ -2\gamma^* t } e^{\gamma^* t'' }\right)_{km}dt'dt''\ . 
\end{split}
\end{equation}

Employing now the definition of ${l}_{i}\left(t\right)$ in Eq.~(\ref{def}), standard properties of Kronecker's delta, and equations~(\ref{R}) and~(\ref{g}), we can write down
\begin{equation}
\label{<ll'>}
\begin{split}
\left<  { l }_{ k }\left( { t' } \right) { l }_{ m }\left( { t'' } \right)  \right>  &=\frac { 1 }{ { \mu  }^{ 2 } } { g }_{ kl }{ g }_{ mn }\left<  { R }_{ k }\left( { t' } \right) { R }_{ m }\left( { t'' } \right)  \right> \\
&=\frac { 2 }{ { \mu  }^{ 2 } } { g }_{ kl }{ g }_{ mn }{ \delta  }_{ ln }\delta \left( t'-t'' \right) \\
&=\frac { 2 }{ { \mu  }^{ 2 } } { g }_{ kl }{ g }_{ ml }\delta \left( t'-t'' \right)\\
&=\frac { 2 }{ { \mu  }^{ 2 } } { D }_{ km }\delta \left( t'-t'' \right) 
 \ .
\end{split}
\end{equation}
We can substitute then Eqs.~(\ref{termino}) and~(\ref{<ll'>}) in Eq.~(\ref{<v2>}) and solve the last integral with the help of $\delta\left(t'-t''\right)$,

\begin{equation}
\label{<v2>def}
\begin{split}
\left<  { v }^{ 2 }\left( t \right)  \right>  &={ { { v }_{ 0 } }_{ j } }{ \left( { e }^{ -2{\gamma}^* t } \right)  }_{ jl }{ { v }_{ 0 } }_{ l }\\
&+{ { { v }_{ 0 } }_{ j } }{ \left( { e }^{ -2{\gamma}^* t } \right)  }_{ jl }\int _{ 0 }^{ t' }{ { \left( { e }^{ {\gamma}^* t'' } \right)  }_{ lm }{ \left<  f\left( { r'' }_{ m } \right)  \right>   }_{ m }dt'' } \\
&+\int _{ 0 }^{ t }{ { \left<  f\left( { r' }_{ k } \right)  \right>   }_{ k }{ \left( { e }^{ {\gamma}^* t' } \right)  }_{ kj }dt' } { \left( { e }^{ -2{\gamma}^* t } \right)  }_{ jl }{ { v }_{ 0 } }_{ l }\\
&+\int_0^t \int_0^{ t' } \left<  f_k(r'_k) f_m(r''_m)  \right>  
\left( e^{\gamma^* t' } e^{ -2\gamma^* t} e^{ \gamma^* t'' }\right)_{km}dt'dt''\\
&+\frac { 1 }{ { \mu  }^{ 2 } } Tr\left[ D{ {\gamma}^*  }^{ -1 }\left( \mathbb{I} -{ e }^{ -2{\gamma}^* t } \right)  \right] 
\end{split}
\end{equation}
(where $\mathbb{I}$ is the identity matrix).

We are now ready to compute the velocity fluctuations ${{\sigma}_v}^2$, 
substituting Eqs.~(\ref{<v>2}) and~(\ref{<v2>def}) in the definition of Eq.~(\ref{sigma}); all terms but one cancel out, except the last of Eq.~(\ref{<v2>}),
\begin{equation}
\label{fluc}
{{\sigma}_v}^2=\frac{1}{{\mu}^{2}}Tr\left[D{{\gamma}^*}^{-1}\left(\mathbb{I}-{e}^{-2{\gamma}^* t}\right)\right] \ .
\end{equation}
Taking the large time limit $t\longrightarrow\infty$ in  Eq.~(\ref{fluc}), 
and undoing the change of variables of Eq.~(\ref{def}), we obtain the fluctuation-dissipation relation quoted in the main text, Eq.~(\ref{fluctuaciones}).

\newpage


\end{document}